\documentclass[useAMS,usenatbib,a4paper]{mn2e}
\usepackage{amssymb,amsmath}
\usepackage{graphicx}
\usepackage{natbib}
\usepackage{epstopdf}
\usepackage{journal_shortcuts}

\def\etal{{\it et al.\ }} 
\def\eg{{\it e.g.\ }} 

\newcommand{\be}{\begin{equation}}  \newcommand{\ba}{\begin{eqnarray}}
\newcommand{\ee}{\end{equation}}  \newcommand{\ea}{\end{eqnarray}}

\newcommand{\Inj}{_{\mathrm{inj}}}

\newcommand{\Turb}{_{\mathrm{turb}}}

\newcommand{\Metal}{_{\mathrm{metal}}}

\newcommand{\Myr}{\,\textrm{Myr}}

\title[Self-Regulation of AGN in Galaxy Clusters]%
{Self-Regulation of AGN in Galaxy Clusters}

\author[M. Br\"uggen \& E. Scannapieco]%
{Marcus Br\"uggen$^1$\footnotemark[1], Evan Scannapieco$^2$
\\
$^1$Jacobs University Bremen, P.O. Box 750\,561, 28725 Bremen,
Germany\\
$^2$School of Earth and Space Exploration,  Arizona State University, P.O.  Box 871404, Tempe, AZ, 85287-1404, USA
}

\begin{document}

\date{Accepted. Received; in original form }

\pagerange{\pageref{firstpage}--\pageref{lastpage}} \pubyear{2009}

\maketitle

\label{firstpage}

\begin{abstract}

Cool cores of galaxy clusters are thought to be heated by low-power active galactic nuclei (AGN), whose accretion is regulated by feedback. However, the interaction between the hot gas ejected by the AGN and the ambient intracluster medium is extremely difficult to simulate as it  involves a wide range of spatial scales and gas that is Rayleigh-Taylor (RT) unstable. Here we present a series of three-dimensional hydrodynamical simulations of a self-regulating AGN in a galaxy cluster.  Our adaptive-mesh simulations include prescriptions for radiative cooling, AGN heating and a subgrid model for RT-driven turbulence, which is crucial to simulate this evolution.  AGN heating is taken to be proportional to the rest-mass energy that is accreted onto the central region of the cluster. For a wide range of feedback efficiencies, the cluster regulates itself for at least several $10^9$ years. Heating balances cooling through a string of outbursts with typical recurrence times of around 80 Myrs, a timescale that depends only on global cluster properties. Under certain conditions we find central dips in the metallicity of the intracluster medium. Provided the sub-grid model used here captures all its key properties, turbulence plays an essential role in the AGN self-regulation in cluster cores. 
\end{abstract}


\section{Introduction}

High-resolution observations from the {\sc Chandra} and {\sc XMM}-Newton X-ray
telescopes have revolutionised our understanding of the hot-intracluster medium  
(ICM).  While a large number of galaxy clusters have strong peaks in 
their central X-ray surface brightness distributions, indicating that their
central gas is cooling rapidly, detailed spectra of such cool-core clusters show 
that this gas fails to cool below $\approx 1$ keV \citep[\eg][]{peterson:01, rafferty:06, mcnamara:07}.
This deficit means that radiative cooling
must be balanced by an unknown energy source. 

Currently, the most successful model for achieving this balance relies
on heating by outflows from an AGN, hosted by the
central cD galaxy \citep[\eg][]{churazov:00,bruggen:02,magliocchetti:07}. 
Galaxies within $0.2r_{200}$ of the centre of groups and clusters 
show a boosted likelihood of hosting large radio-loud
jets of AGN-driven material \citep{burns:90,best:05}. The energies 
from such outbursts are
comparable to those needed to stop the gas from cooling \citep[\eg][]{simionescu:08}, and the mean
power of the outbursts is well-correlated with the radiated power \citep{birzan:04}.
Furthermore,  AGN feedback from the central cD increases in
proportion to the cooling  luminosity, as expected in an operational
feedback loop \citep[\eg][]{birzan:04,rafferty:06}.

Yet despite these many clues,
the nature of the feedback mechanism is still not understood. In clusters,
energy coming from a region less than a parsec across has to regulate
a volume of gas that is about a few hundred kpc across.  This regulation
must maintain the balance between heating and cooling for a long time,
at least since a redshift of $z \approx 0.4$ \citep{bauer:05}. Moreover,
it must hold in systems with luminosities that span about three orders
of magnitude \citep{churazov:05}. Finally, the output by the AGN must be regulated by the physical conditions in the vicinity of the AGN, rather than the properties of the host
galaxy in which it is contained \citep{best:07}.

In general, gas can make its way onto an AGN disc directly from the  hot 
gas via Bondi-Hoyle accretion \citep[\eg][]{allen:06} or in the form of cold blobs 
\citep[\eg][]{soker:06}. It is likely that the AGN feedback  
in clusters is caused by accretion of hot gas onto pre-formed massive black
holes in old elliptical galaxies  \citep[\eg][]{best:07, hardcastle:07}, which
produce low-power radio jets but
relatively little emission at optical, ultraviolet and infrared
wavelengths.  On the other hand, the fuelling by cold gas may be responsible for AGN with strong emission lines \citep[\eg][]{hardcastle:07}  but the supply of powerful high-excitation sources does not have a direct connection to the hot phase. Most efforts to explain feedback have concentrated on the
hot-accretion scenario. 

 
\cite{churazov:02} considered a toy model of ICM feedback regulated by hot gas 
accretion, in which black hole growth was set by the classic Bondi formula, with
$\dot M \propto n_e T^{-3/2}$.  For a ratio of specific heats
of 5/3, this rate only depends on the entropy and is
directly affected by heating and cooling.  The ICM in the model
responded  to the  heating by expanding, which lowered the gas density and
accretion rate. As the gas radiated away energy, the atmosphere
contracted and the accretion rate rose again. Application of the results to
the Virgo cluster suggested that accretion proceeded at approximately the
Bondi rate down to a few gravitational radii and that most of the
power was carried away by an outflow. Comparison of the accretion rate
predicted by the Bondi formula with the luminosity of the cooling flow
region in M87 suggested a high efficiency for transforming the rest
mass of accreted material into mechanical power of an outflow at the
level of a few per cent.  Recently, \cite{allen:06} found evidence that the power of 
outbursts in giant elliptical galaxies scales with the Bondi accretion rate.  

However, the full simulation of this feedback cycle remains elusive.
Direct simulation of a galaxy cluster from Mpc scales down to the parsec scale
at which mass accretes onto the central supermassive black hole
is still out of reach even for the latest generation of adaptive-mesh refinement
(AMR) codes. Hence, in most 
hydrodynamical simulations of AGN feedback, the energy input into the ICM is set by
hand, rather than computed from the conditions surrounding the 
central galaxy.  Only \cite{vernaleo:06}, \cite{brighenti:06}
and \cite{cattaneo:07} attempted hydrodynamic simulations of self-regulated
accretion, and none of these studies was able to reproduce 
all the observed properties of cool-core clusters.

\cite{vernaleo:06} carried out a set of three-dimensional cluster simulations
in which they took the kinetic energy of
the AGN jet to be  proportional to the instantaneous mass accretion rate
across an inner spherical boundary.
They found that their simulations were incapable of balancing heating
and cooling as the jet formed a low-density channel, which stopped
the hot gas from coupling effectively to the surrounding ICM. A delay between the mass accretion rate and
the response of the jet was introduced in further simulations, but could not alleviate this problem. However,
\cite{heinz:06} have shown that in more realistic initial conditions
the ambient bulk motions of the ICM prevent the formation of such
channels, though their simulations were not self-regulating.

\cite{brighenti:06} simulated self-regulated mass-loaded jets
that resemble disk winds. They presented two-dimensional
hydrodynamical models where outflows were generated by
assigning a fixed velocity to gas that flowed into a biconical source
region at the cluster centre. The acceleration of gas in the source
cones was triggered when gas flowed into
the innermost zones.  As the jet proceeded, it entrained additional
ambient gas and its mass flux increased. After many gigayears, the
time-averaged gas temperature profile resembled observations and very little gas cooled below the virial temperature.

However, feedback is at least partly mediated by low-power Fanaroff-Riley type I (FR I) sources and this has not been simulated before. 
\cite{cattaneo:07} carried out three-dimensional simulations
in which AGN jets were regulated via a Bondi-accretion
rate measured at the cluster centre starting from hydrostatic initial
conditions. For their setup, a very violent AGN phase
($L\sim10^{45}$ erg/s) occurred lasting for about 250 million years,
followed by a period of at least 4 gigayears during which the
accretion rate of the black hole and the mass of the cold gas in the
cluster core stayed constant with a luminosity of a few times $10^{43}$
erg s$^{-1}$. This scenario does not explain the recurrent, low-power
FR I sources that are observed in local cool-core
clusters.

The biggest obstacle to these simulations is that it is still unclear
how the mechanical power of the AGN is transformed into thermal energy
and distributed throughout the cluster.
\cite{nulsen:07} pointed out that theoretically one would expect 
the energy to be dissipated locally.  
As the bubble rises, they argue,
the ICM falls inward around it to fill the space  it
occupied, converting  gravitational potential energy to kinetic
energy.  Details of the flow
depend on the viscosity, which is poorly determined. If it is high,
the flow is laminar and the kinetic energy is dissipated as heat over
a region  comparable in size to the cavity. If the viscosity is 
low, the Reynolds number is high and the flow would form
a turbulent region of a similar size to the cavity.
Turbulent kinetic energy is dissipated in the turnover time of the
largest  eddies, so that the dissipation time $t_{\rm diss} \approx
r_b/v_{\rm turb}$, where $r_b$ is the radius  of the bubble and
$v_{\rm turb}$ is the speed of the eddy, which is comparable to the
speed of the cavity. Since $t_{\rm diss} v_b \approx r_b$, much of the
turbulent kinetic energy  is dissipated in a region of comparable size
to the cavity. Thus, regardless of  the viscosity, the kinetic energy
created by cavity motion is deposited locally.  
Other authors have pointed out the importance of sound waves and weak
shocks in dissipating heat \citep{fabian:03, bruggen:07, sanders:07,
sanders:08}, and considered  the possibility that AGN bubbles are
stabilized by magnetic fields (De Young 2003; Br\" uggen \& Kaiser
2001; Ruszkowski \etal 2007) or massive jets \citep{sternberg:08}.

In \cite{scannapieco:08} (SB08 hereafter),  we showed that even in the
absence of magnetic effects, the evolution of AGN bubbles
cannot presently be captured reliably by direct simulations.
While pure hydrodynamic simulations indicate that AGN bubbles are disrupted into
resolution-dependent pockets of  underdense gas, proper modeling of
subgrid turbulence indicates that this a poor approximation to a
turbulent cascade that continues far beyond current resolution limits.
Adding a subgrid model of turbulence and mixing \citep{dimonte:06}  to
the adaptive-mesh hydrodynamic code, FLASH3, we found that
Rayleigh-Taylor (RT) instabilities act to effectively mix the heated region
with its surroundings, while at the same time preserving it as a
coherent structure, consistent with observations.
\cite{cattaneo:07} also pointed out the importance of
turbulence in mediating the jet power to the accretion mechanism,
although they were unable to resolve it.

In this paper, we explore how a subgrid model for turbulence provides
the missing piece necessary to construct a full numerical 
simulation of a self-regulating jet at the centre of cool-core clusters.
Furthermore, with no fine tuning, such simulations naturally explain both
the mechanical power of AGN, and their duty cycles.
The structure of the paper is as follows: In \S2, we describe our
method including the algorithm and the setup, and in \S3 we present our
results.  A discussion is given in  \S4.

\section{Method}

\subsection{Simulation and Turbulence Modeling}

Out simulations were performed with  FLASH version 3.0 \citep{fryxell:00}, a
multidimensional adaptive mesh refinement hydrodynamics code. It
solves the Riemann problem on a Cartesian grid using a
directionally-split  Piecewise-Parabolic Method (PPM) solver
\citep{colella:84}.
While the direct simulation of turbulence is extremely challenging,
computationally expensive, and dependent on resolution  
\citep[\eg][]{glimm:01}, its behaviour can be approximated to a good degree of
accuracy by adopting a sub-grid approach. Recently, \cite{dimonte:06}, described a sub-grid 
model that is especially suited to capturing the buoyancy-driven, turbulent evolution
of AGN bubbles. The model captures the self-similar growth of the RT
and Richtmyer-Meshkov (RM) instabilities by augmenting the mean hydrodynamics equations
with evolution equations for the turbulent kinetic energy per unit
mass and the scale length of the dominant eddies. The
equations are based on buoyancy-drag models for RT and RM flows, but
constructed with local parameters so that they can be applied to
multi-dimensional flows with multiple materials.  The model is
self-similar, conserves energy, preserves Galilean invariance, and
works in the presence of shocks. 

The mean flow fluid equations in this case are given by 
\ba
\frac{\partial {\bar \rho}}{\partial t} + \frac{\partial {\bar \rho}
\tilde u_j}{\partial x_j} &=& 0,
\label{eq:rho}\\
\frac{\partial {\bar \rho} \tilde u_i}{\partial t} + \frac{\partial
{\bar \rho} \tilde u_i \tilde u_j}{\partial x_j} 
+\frac{\partial
P}{\partial x_i} +   \rho \frac{\partial \Phi}{\partial x_i} &=&  C_P \frac{\bar \rho K}{\partial x_i},
\label{eq:u}\\
\frac{\partial {\bar \rho} E}{\partial t} + \frac{\partial {\bar \rho}
E \tilde u_j}{\partial x_j} 
+\frac{\partial P \tilde u_j}{\partial x_j} 
&=&  \frac{\partial}{\partial x_j} \left(
\frac{\mu_t}{N_E}  \frac{\partial E}{\partial x_j}\right)
- S_K, \nonumber
\label{eq:E} 
\ea 
where $t$ and ${\bf x}$ are time and position variables, $\bar
\rho({\bf x},t)$ is the average density field,  $\tilde u_i({\bf
x},t)$  is the mass-averaged mean-flow velocity field in the $i$
direction, $P({\bf x},t)$ is the mean pressure, $\Phi({\bf x}, t)$ is the gravitational potential, 
and $E({\bf x},t)$ is the mean
internal energy per unit mass. The terms on the right hand side of
these equations couple the subgrid quantities to the mean flow.
These three terms model: (i) the excess ``pressure'' arising from the  turbulent kinetic energy 
per unit mass $K,$ which is scaled by a constant $C_P$; (ii) turbulent
mixing, which is modeled as a turbulent viscosity $\mu_t$, scaled by a constant $N_E$; and
(iii) an explicit source term, $S_K$ that 
contains both RT and RM contributions,
and removes energy from the mean flow and moves it into the turbulent field.
This is given by
\be 
S_K = \bar \rho V \left[ C_B A_i g_i - C_D \frac{V^2}{L} \right],
\label{eq:sk}
\ee 
where $V \equiv \sqrt{2 K}$ is the average turbulent velocity, $L$ is the turbulent
eddy scale, $g_i \equiv - (1/\rho) \partial P/\partial
x_i$ is the gravitational acceleration, the coefficients $C_B$ and $C_D$ are fit to experiments ($C_B=0.84$ and $C_D=1.25$) and $A_i$ is the Atwood number in the $i$ direction. In the code, we  determine this as
\be
 A_i = \frac{\bar \rho_+ - \bar \rho_-}{\bar \rho_+ +
\bar \rho_-} + C_A \frac{L}{\bar \rho + L |\partial \bar \rho/\partial
x_i|}  \frac{ \partial \bar \rho}{\partial x_i},
\label{eq:Ai}
\ee 
where $C_A $ is a constant and $\bar \rho_-$ and $\bar \rho_+$ are the densities on
the back and front boundaries of the cell in the $i$ direction.
This source term causes turbulence to decay away at a characteristic time scale of 
$\approx L/V$ in the absence of external driving,
while the growth term causes turbulent velocity to grow at a rate $\dot V \approx g,$
whenever the gravitational acceleration opposes the density gradient.

The turbulence quantities that appear in these equations are
calculated from evolution equations for $L$ and $K$.  The eddy scale
$L$ must be evolved because the buoyancy-driven RT and RM
instabilities depend on the eddy size, which is expected to grow
self-similarly. Simple equations that include
diffusion, production, and compression are given by
\be 
\frac{\partial \bar \rho L}{\partial t} + \frac{\partial \bar
\rho L \tilde u_j}{\partial x_j} = \frac{\partial}{\partial x_j}
\left(  \frac{\mu_t}{N_L}  \frac{\partial L}{\partial x_j}\right) +
\bar \rho V + C_C \bar \rho L \frac{\partial \tilde u_i}{\partial x_i}
\label{eq:L},
\ee 
and 
\be 
\frac{\partial \bar \rho K}{\partial t} + \frac{\partial
\bar \rho K \tilde u_j}{\partial x_j} = \frac{\partial}{\partial x_j}
\left(  \frac{\mu_t}{N_K}  \frac{\partial K}{\partial x_j}\right) -
C_P \rho K \frac{\partial \tilde u_i}{\partial x_i} + S_K,
\label{eq:K}
\ee 
where $N_K$, $N_L$, $C_C,$ and $C_P$ are constants ($N_K=1.0$, $N_L=0.5$, $C_C=1/3$ and $C_P=2/3$). See \cite{dimonte:06} for details on how these constants are determined.
In the $L$ equation the three terms on the right hand side of the
equation represent, respectively: turbulent diffusion, growth of
eddies through turbulent motions, and growth of eddies due to motions
in the mean flow.  In the $K$ equation the three terms on the right
hand side represent, respectively: turbulent diffusion, the work
associated with the turbulent stress, and the source term $S_K,$
which also appears in the internal energy equation to conserve energy.
Finally, the turbulent viscosity is calculated as  
\be 
\mu_t = C_\mu \bar \rho L \sqrt{2 K},
\label{eq:mut}
\ee where $C_\mu$ is a constant.  Further details of this model and
our implementation are given in SB08.

All our simulations are performed in a cubic three-dimensional region
with reflecting boundaries that is 680 kpc on a side,  although the cool-core region
in which we are most interested lies within $\lesssim 100$ kpc of the center.   
For our grid, we chose a block size of $8^3$ zones and an  unrefined root grid with
$8^3$ blocks, for a native resolution of 10.6 kpc.   The refinement
criteria are the standard density and pressure criteria, and we allow
for 4  levels of refinement beyond the base grid, corresponding to  a
minimum cell size of 0.66 kpc, and an effective grid of 1024$^3$
zones.  In all zones $K$ is initialized to $10^{-10} (p/\rho)/2.$ and
$L$ is initialized to $10^{-5} \sqrt{p/\rho} \sqrt{1/G \rho}.$

\subsection{Cluster Profile}

For our overall cluster profile, we adopted the model
described in \cite{roediger:07}, which was constructed to reproduce the properties of the
brightest X-ray cluster A426 (Perseus) that has been studied
extensively with {\sc Chandra} and {\sc XMM}-Newton.
In this case, the electron density $n_{\rm e}$ and temperature $T_{\rm e}$ 
profiles are based on the deprojected XMM-Newton data \citep{churazov:03}
which are in broad agreement with the {\sc Chandra} data \citep{schmidt:02, sanders:04}; namely:
\be
n_{\rm e}=\frac{4.6\times10^{-2}}{[1+(\frac{r}{57})^2]^{1.8}}+
\frac{4.8\times10^{-3}}{[1+(\frac{r}{200})^2]^{0.87}}~~~{\rm cm}^{-3},
\label{eq:ne}
\ee
and
\be
T_{\rm e}=7\times\frac{[1+(\frac{r}{71})^3]}
    {[2.3+(\frac{r}{71})^3]}~~~{\rm keV},
\label{eq:te}
\ee
where $r$ is measured in kpc. Furthermore, the hydrogen number density was 
assumed to be related to the electron number density as $n_{H}=n_{\rm e}/1.2$.
The static, spherically-symmetric gravitational potential, $\Phi(R),$  was set such that 
the cluster was in hydrostatic equilibrium.

As in Scannapieco \& Br\" uggen (2008), we computed the radiative losses in
each cell in the optically-thin limit, with an emissivity $\epsilon =  \Lambda(T) n_H n_e$
where $n_H$ and $n_e$ are the number density of hydrogen and electrons, respectively, and
the cooling function $\Lambda (T )$ describes radiative losses from the optically thin plasma, as in \cite{sarazin:86} \citep[see also][]{raymond:76, peres:82}.

The metal injection rate in the central galaxy was assumed to be
proportional to the light distribution, and modeled with a
Hernquist profile given by
\begin{equation}
\dot\rho\Metal (r,t) = \frac{\dot M\Metal{}_0} {2\pi} \, 
\frac{a}{r} \, \frac{1}{(r+a)^3},
\label{eq:hernquist}
\end{equation}
with a scale radius of $a= 10$ kpc and a total metal injection rate of $\dot M\Metal{}_0  =  9.5\times 10^4 \,M_\odot\Myr^{-1}.$ This is the integral mass injection rate for all elements heavier than helium. The sources of metals are supernovae Ia and stellar mass loss \citep{rebusco:05}. 

In order to be able to trace the metal distribution, we utilize
a mass scalar to represent the local metal fraction in each
cell, $F \equiv \rho\Metal/\bar \rho.$
Hence, the quantity $F \bar \rho$ gives the local metal density, $\rho\Metal$,
which has a continuity equation including the metal source, given by
\be
\frac{\partial {\bar \rho} F}{\partial t} +
\frac{\partial {\bar \rho} F \tilde u_j}{\partial x_j} =
\frac{\partial}{\partial x_j}
\left(  \frac{\mu_t}{N_F} 
\frac{\partial F}{\partial x_j}\right)+ \frac{\dot\rho\Metal}{\bar \rho}
\label{eq:continuity_metals}
\ee
where $N_F$ is a dimensionless constant.
Furthermore, we assumed that
the metal fraction is small at all times. Hence, we could neglect
$\dot\rho\Metal$ as a source term in the continuity equation for the gas
density. Also, the mass fraction can be scaled to any value in post-processing.

\subsection{AGN Feedback} 

Bubbles in the ICM are thought to be inflated by a pair of ambipolar
jets from the central AGN that inject energy into small
regions at their terminal points, which expand until they reach
pressure equilibrium with the surrounding medium
\citep{blandford:74}. The result is a pair of underdense, hot bubbles
on opposite sides of the cluster centre. In order to produce such bubbles, we 
injected  energy  into two small spheres of radius $r_{\rm bubble},$ each a distance 
$r_{\rm bubble}$ from the centre.

Unlike in our previous simulations,
the energy input rate was not predetermined, but instead calculated from the instantaneous
conditions near the centre of the cluster.   In particular, we considered a model in which a 
fixed fraction of the gas within the central 3 kpc of the cluster accretes onto the
central supermassive black hole within a cooling time, and a fixed fraction of the
$mc^2$  rest mass energy of this accreted gas is returned to the ICM by increasing the pressure
within the bubble regions.   This is similar to the approach
taken by \cite{vernaleo:06}, who calculated AGN feedback based on the mass inflow rate along the
{\rm 10 kpc} inner boundary of a spherical grid.  Here our choice of the volume within $3$ kpc of the cluster
centre is taken to sample the conditions as near to the central supermassive black hole as possible, while still 
averaging over enough cells to avoid numerical effects. 

The  efficiency of gas accretion in powering the bubbles
was chosen by scaling to typical numbers observed in galaxy clusters.  We took the rate of change of the 
pressure within the bubble regions to be given by 
\be
\frac{3}{2} \dot P \Inj(t)  \frac{4 \pi}{3} r_{\rm bubble}^3 = 
 E_{\rm bubble} \frac{75 {\rm Myrs}} {t_{\rm cool, 3 kpc}(t)}
        \frac{ M_{\rm gas, 3 kpc}(t)}{M_{\rm gas,3 kpc}(0)} ,
\ee
where $t_{\rm cool, 3kpc} (t)$ is calculated by dividing the total internal energy of the gas within the central
3 kpc by the total radiation rate in this regions, and $M_{\rm gas, 3 kpc}(t)$ is the total gas within the central 3 kpc, and
$E_{\rm bubble} \approx 5 \times 10^{59}$ ergs corresponds to the typical energies necessary to inflate the X-ray cavities 
observed in cool-core clusters \citep[\eg][]{forman:07}, which are generated approximately every 75 Myrs. 
This prescription corresponds to an overall efficiency of $3.2 \times 10^{-3} 
E_{\rm bubble}/10^{60} {\rm ergs}$ of the  $m c^2$ energy of the mass cooling rate within the central 3 kpc. 
Note that choosing a typical value of  $E_{\rm bubble} \approx 5 \times 10^{59}$ ergs would mean that our energy input is only about
$4 \times 10^{-5}$ of the rest mass energy of the mass cooling rate within the central $10$ kpc, and thus our models generally 
have energy input levels similar to that considered by Vernaleo and Reynolds (2006).  

Finally, we started all runs with an initial pair of bubbles overpressured by a factor of 
\be
 P_{\rm bubble}(0)  \frac{4 \pi}{3} r_{\rm bubble}^3 = E_{\rm bubble},
\ee
so as to create some initial ``seed'' turbulence, to help increasing mixing at early times and
speed the simulations towards their eventual quasi-steady states. 
The choices of $E_{\rm bubble}$ and $r_{\rm bubble}$ for each of the runs carried out in this study are given in Table \ref{tab:runs}.

\begin{table}
\caption{Run Parameters}
\label{tab:runs}
\centering\begin{tabular}{llll} \\ 
\hline  Run    & Subgrid & $E_{\rm bubble}$ &  $r_{\rm bubble}$\\  
 Name   & Model   & (ergs)           & (kpc)      \\     
\hline  
N5-10  & No  & $5 \times 10^{59}$  & 10 \\ 
D5-10  & Yes & $5 \times 10^{59}$  & 10 \\
D5-8   & Yes & $5 \times 10^{59}$  & 8  \\ 
D5-12  & Yes & $5 \times 10^{59}$  & 12 \\ 
D2-10  & Yes & $2 \times 10^{59}$  & 10 \\ 
D10-10 & Yes & $1 \times 10^{60}$  & 10 \\ 
D20-10 & Yes & $2 \times 10^{60}$  & 10 \\ 

\hline
\end{tabular}
\end{table}

\section{Results}

\subsection{Self-regulating Oscillations}

\begin{figure*}
\includegraphics[trim=0 0 0 0,clip,width=\textwidth]{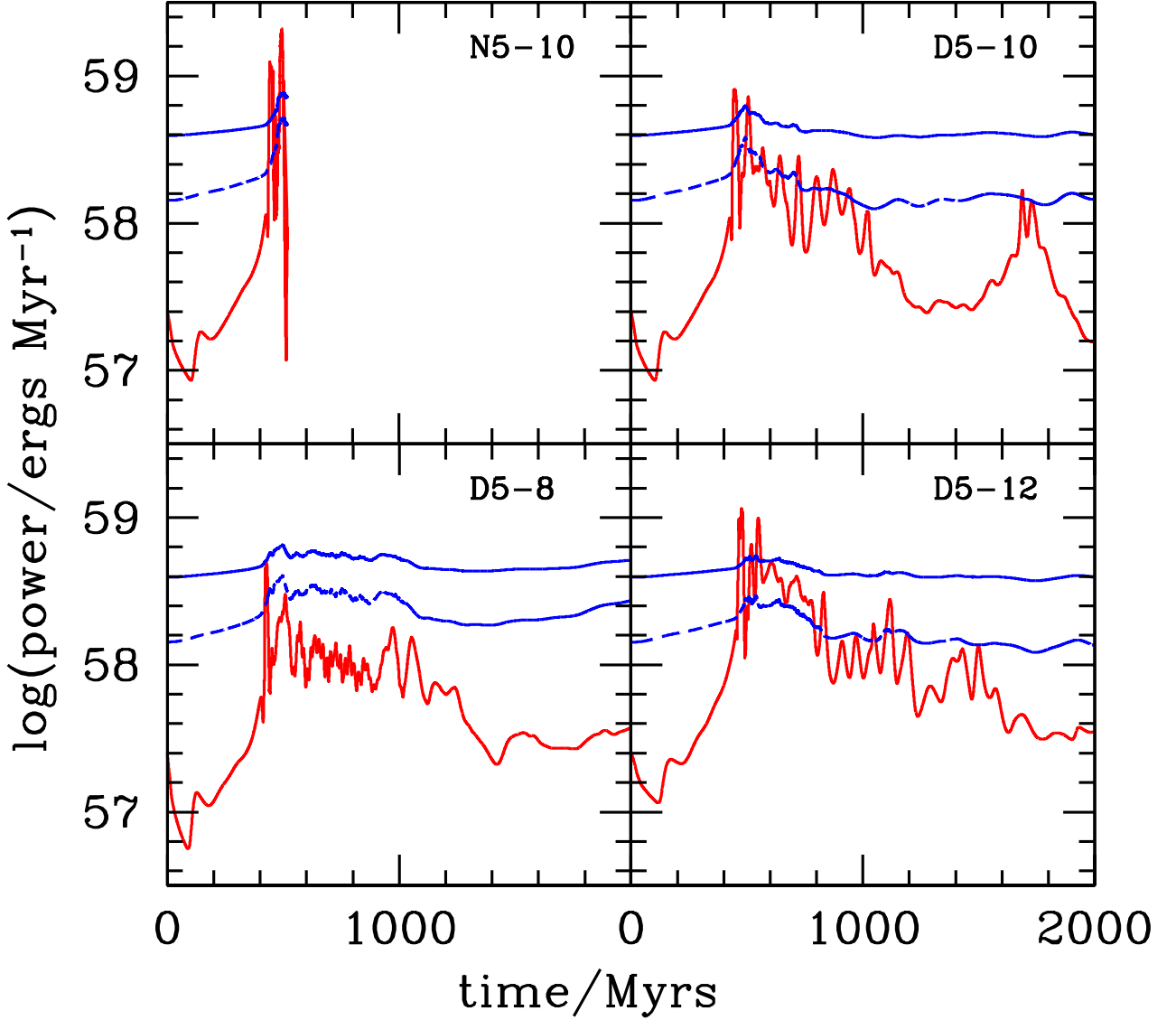}

\caption{Bubble power and integrated cooling rates for D5-10 (top
left), N5-10 (top right), D5-8 (bottom left) and D5-12 (bottom
right). The solid blue lines shows the cooling losses in the entire
simulation volume, the dashed blue line show the cooling losses in the
central 100 kpc of the cluster and the red line shows the energy put
into bubbles.  The lines for our fiducial run without a subgrid
turbulence model   (named N5-10) stop at about 450 million year
because this is where catastrophic cooling in the centre halted the
simulation.}
\label{fig:power1}
\end{figure*}

\begin{figure*}
\includegraphics[trim=0 0 0 0,clip,width=0.8\textwidth]{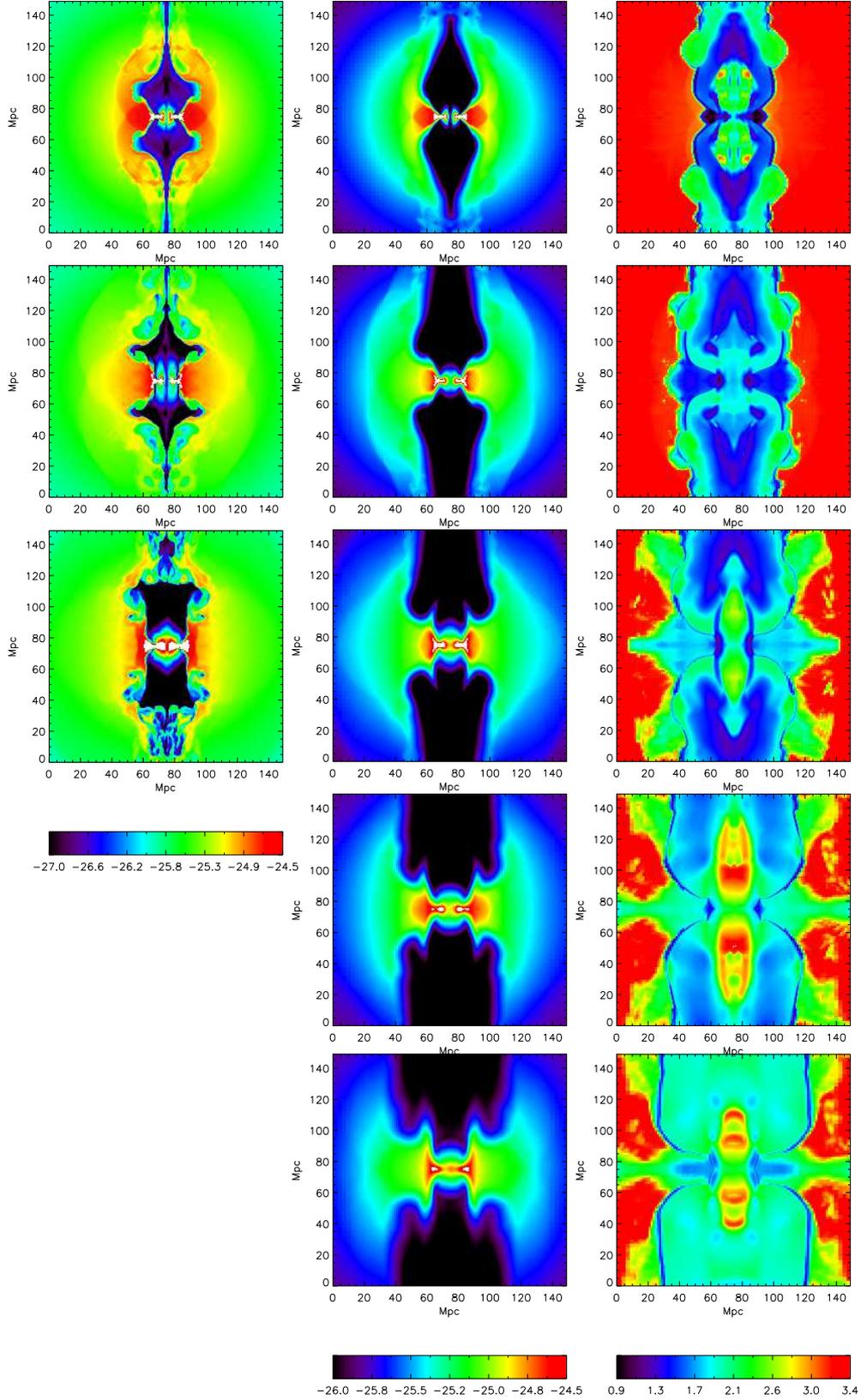}
\caption{Slices of  $\log(\rho/ {\rm g \, cm^{-3}})$ (1st column for N5-10 and 2nd column for D5-10) and $\log$ turbulent diffusion time in Myrs (3rd column for D5-10) for times near to when the non-turbulent run cools catastrophically. From top to bottom $t=$ 460, 480, 500, 520 and 540 Myrs.}
\label{fig:dens_tdiff23}
\end{figure*} 

\begin{figure*}
\includegraphics[trim=0 0 0 0,clip,width=0.8\textwidth]{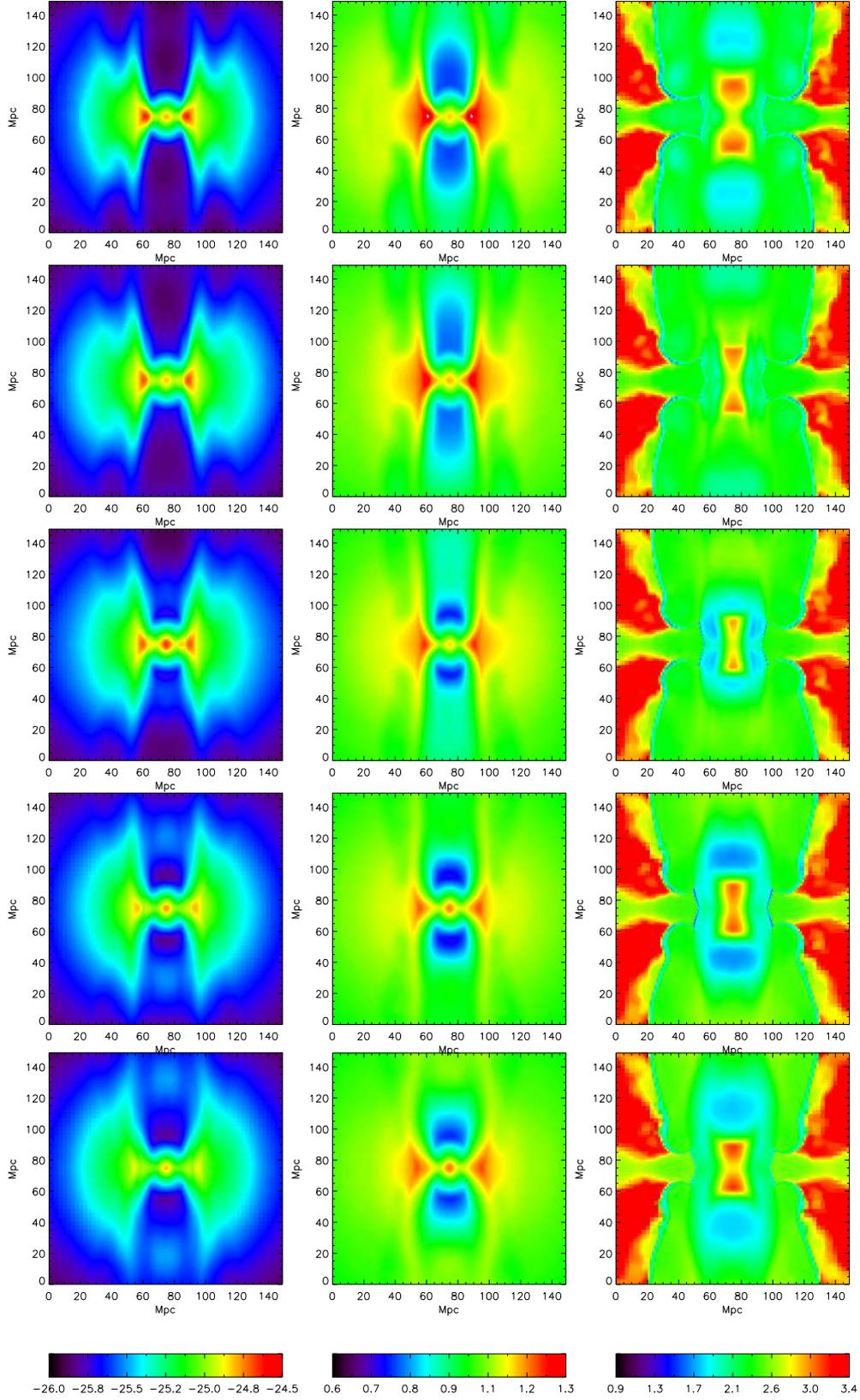}

\caption{Slices of $\log(\rho/ {\rm g \, cm^{-3}})$ (1st column), $\log$ sound crossing time in Myrs (2nd column), and  $\log$ turbulent diffusion time in Myrs (3rd column), 
as the fiducial subgrid run (D5-10) executes a full  $\approx 80$ Myr cycle as described in the text.
From top to bottom $t=$ 600, 620, 640, 660 and 680 Myrs (all from D5-10).}
\label{fig:dens_mut}
\end{figure*} 

\begin{figure*}
\includegraphics[trim=0 0 0 0,clip,width=\textwidth]{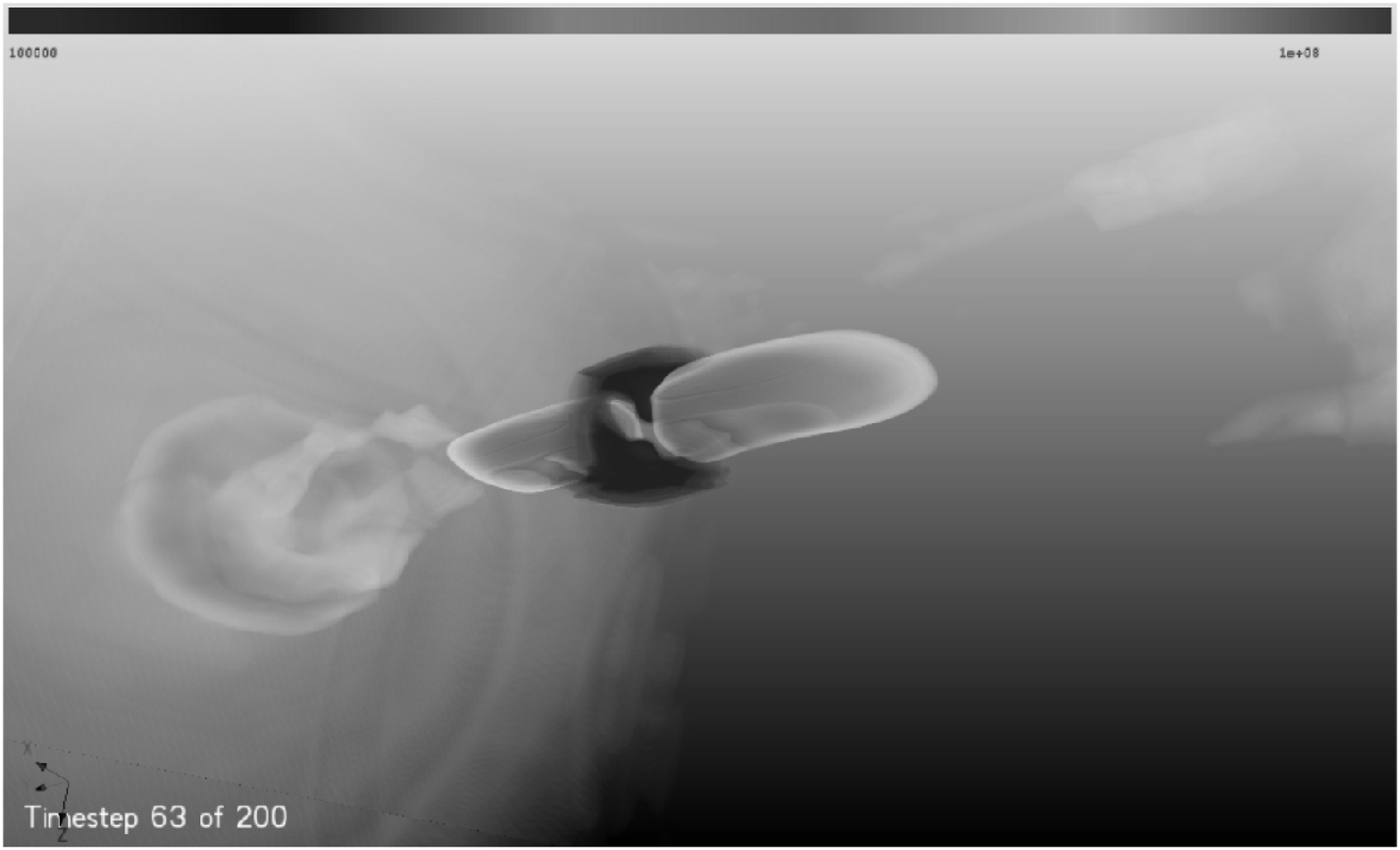}
\caption{Volume rendering of the temperature in our fiducial run D5-10 at t= 630 Myrs. The blue ring shows the cool gas accreting onto the centre; the red and yellow blobs represent the hot gas ejected by the AGN. On the left side one can see an older bubble from an earlier outburst.}
\label{fig:64}
\end{figure*} 

As a control case, we first present the results of a model without 
subgrid turbulence (5N-10) in which we have chosen
$E_{\rm bubble} \approx 5\times 10^{59}$ ergs and $r_{\rm bubble} = 10$ kpc.
The evolution of the cooling and AGN heating in 
this case is summarised in the top left panel in Fig.~\ref{fig:power1}. 
Here we see that the bubbles do not couple to the central region that determines
the activity of the AGN. Instead, cool gas accumulates in the centre, increasing
the AGN output as shown in this figure, as well as leading to the cold dense
central  regions shown in Fig.~\ref{fig:dens_tdiff23}.
This cooling eventually leads to a drastic outburst of the AGN, which is
apparent in a sharp upturn in the heating rate in Fig.~\ref{fig:power1},
and the large empty cavities in the third row in
Fig.~\ref{fig:dens_tdiff23}.  However this heating does not stop the flow of
gas onto the AGN along the midplane.  Instead
cooling increases catastrophically and we are forced to stop the simulation. 

\begin{figure*}
\includegraphics[trim=0 0 0 0,clip,width=\textwidth]{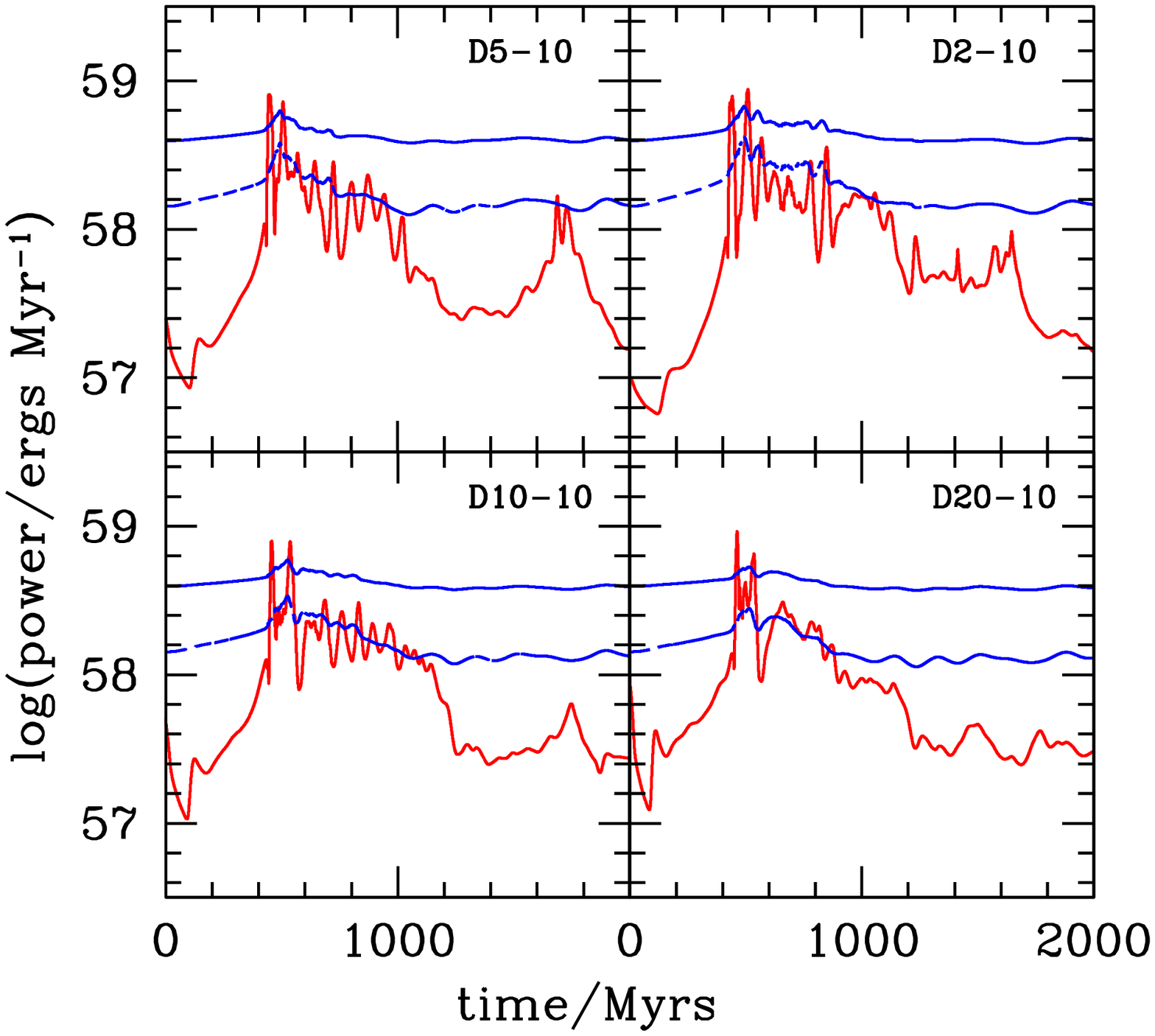}
\caption{Bubble power and integrated cooling rates for D5-10 (top left), D2-10 (top right), D10-10 (bottom left) and D20-10 (bottom right). Meaning of lines as in Fig. 1.}
\label{fig:power2}
\end{figure*} 

On the other hand, when the subgrid model for the turbulence is
switched on, as in run 5D-10, the bubbles couple much more
effectively to the central region.
  As seen in the second and third
columns of  Fig.~\ref{fig:dens_tdiff23}, the cold gas that forms in a
ring around the ICM, shown in white, is slowly eroded and the AGN
activity subsides. The turbulent diffusion or turn-over time, $L/V,$ is plotted in
the third column of Fig.~\ref{fig:dens_tdiff23}. It shows that near the ring of cool gas in the centre of the cluster this time is of the order of 80 Myrs.
This turbulence mixes in hot material near the centre of the cluster,
heating the cool inflowing ring of intracluster gas and stopping the AGN outburst.

Thus instead of cooling catastrophically, the cluster in the simulation with subgrid turbulence begins to execute a
series of oscillations as indicated in Fig. \ref{fig:dens_mut}.   The top row of this plot shows the
configuration at 600 Myrs, shortly after an outburst of AGN activity.  From this time to 640 Myrs,
 turbulence decays away at the turn-over time scale, $L/V,$ and mixing near the cluster center becomes progressively less
efficient. This leads to an increased level of accretion near the core, as more and more cold gas makes its way onto the AGN.  The result is a new burst of AGN activity, which drives a bubble on the order of a sound crossing time. The RT-unstable bubble leads to a rise of the turbulent viscosity and mixing, which quickly quenches  accretion when the turbulent length scale grows to be of the order of the scale of the accretion flow.   At this point the AGN remains relatively quiescent until the turbulence decays again on a time of order $L/V$, repeating the cycle as shown in the bottom panels of Fig.\ \ref{fig:dens_mut}.
This interaction of the AGN-heated regions with the cool inflowing gas is also
illustrated in the volume-rendered image shown in Fig.\ \ref{fig:64}.

It is important to point out that we did not tweak the parameters of the subgrid
model to achieve the self-regulating cycle. Instead, these parameters
are determined by experimental studies of the RT instability and
are the same as those employed in \cite{scannapieco:08}.
Note also that the time scale for this cycle 
is not the sound crossing time $l/c_s$ for
the central region, which is of the order of 10 Myrs.
Instead the period between the episodes of AGN
fueling is determined by the time it takes for the turbulence to 
decay away after it has begun to mix the cold gas in the cluster centre.  
This is given by
\be  
t_{\rm duty} \approx l/v\Turb ,
\label{eq:dutycycle1}
\ee 
where $l$ is the distance between the accretion region of the AGN and the 
heating region.  In our fiducial simulation, $l=r_{\rm bubble}=10$ kpc and is clearly resolved. $v\Turb$ is a typical turbulent velocity, which we assume grows in a dynamical time given by $v\Turb \sim gt \sim g l/c_s$, and $c_s$ is 
the sound speed. Because the cluster is in hydrostatic balance, the gravitational acceleration, $g$ can be written as 
\be  
g = c_s^2\frac{1}{\rho}\frac{d\rho}{dr} \equiv c_s^2/r_0 .
\label{eq:g}
\ee
This leads to the turbulent velocity  
\be  v\Turb = g l/c_s \approx c_s \frac{l}{r_0},
\label{eq:vt}
\ee  and
\be  
t_{\rm duty} \approx \frac{r_0}{c_s}.
\label{eq:dutycycle2}
\ee 
According to this estimate the turbulent speed is given by $c_s l /  
r_0$, which is roughly 700 km s$^{-1}$ (10 kpc/ 60 kpc) $\sim 100$ km s 
$^{-1}$, which is indeed what we find in our simulations.
This means that the size of the bubbles 
does not enter the expression for the duty cycle as would be
expected in a cycle regulated by laminar flow.   It is the
properties of the cluster itself, rather than the jet physics
of the central radio source that are setting the recurrence time
of the bubbles. Thus the size of the bubbles themselves should not
affect the recurrence time of the AGN that we
see in Fig.~\ref{fig:power1} - \ref{fig:dens_mut}.
For the cluster simulated here, the central sound speed is around 700
km s$^{-1}$, the scale height of the cluster, $r_0$, is
around 60 kpc so the duty cycle is 60 kpc / 700 km s$^{-1}
\approx 80$ Myrs.

To test this hypothesis we have varied the geometry of the injection region, leading
to the heating and cooling evolution shown in the lower panels of 
Fig.~\ref{fig:power1}.  As in run D5-10,
the time between two subsequent outbursts for runs D5-8 and D5-12
is roughly 80 Myrs.   
Also, in all cases the activity goes down when
the cool core has been destroyed by recurrent AGN activity.
Again, this large overall time scale, which is $\approx 1$ Gyr,  is not dependent on 
the particulars of bubble injection.  Rather it is 
the time it takes for the turbulence to dissipate through
the cool core of the clusters, which is given by
\be  
t_{\rm cc}  \approx R_{\rm cc} / v\Turb \approx 100\, {\rm kpc}/ 100\,
{\rm km/s} \approx 1\, {\rm Gyr } \ .
\label{eq:tdiff}
\ee
After this the core has been heated sufficiently to stop the gas
inflow. It then takes a central cooling time of about 1 Gyr for the
cool core to reform and the intermittent AGN activity to resume.

\begin{figure*}
\includegraphics[trim=0 0 0 0,clip,width=\textwidth]{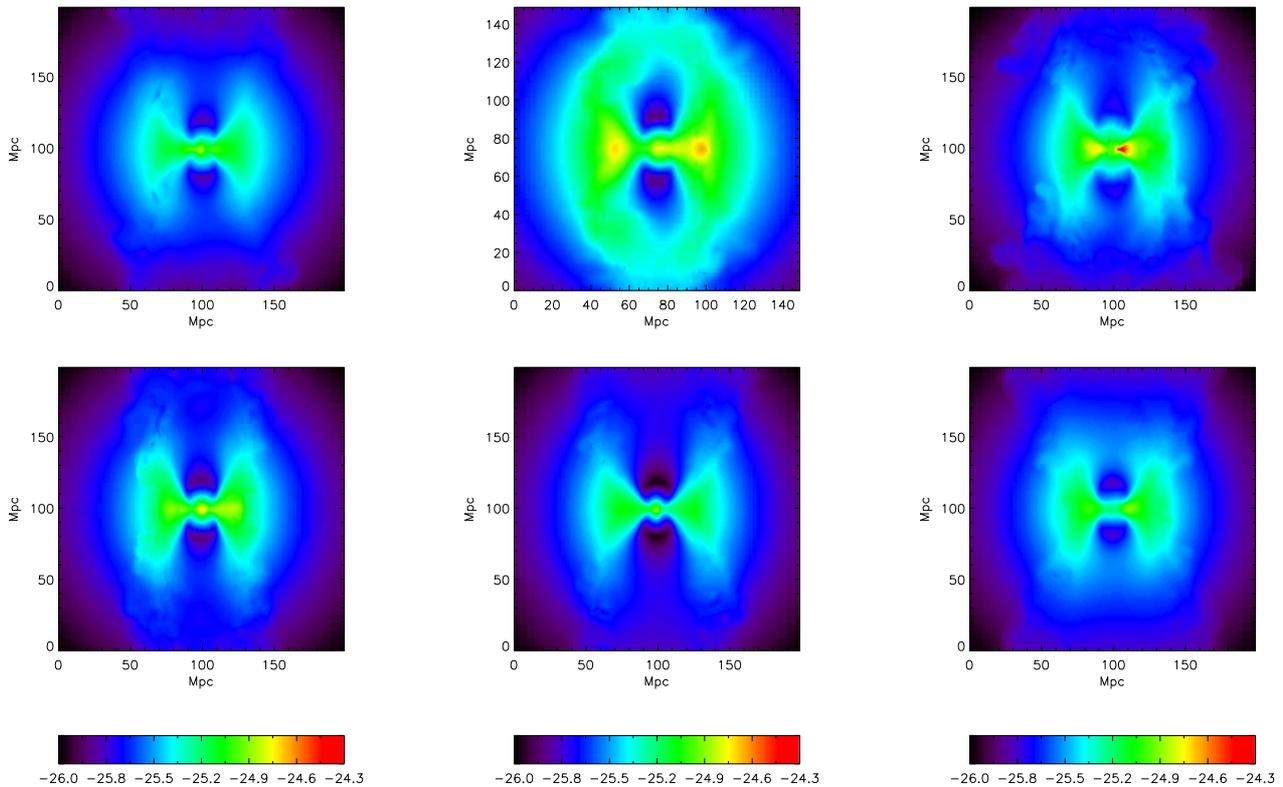}
\caption{Slices of the logarithm of the density at 1 Gyr. Top from left to right: D5-10, D5-8, D5-12 Bottom from left to right: D2-10, D10-10, D20-10.}
\label{fig:slices50}
\end{figure*} 

\begin{figure*}
\includegraphics[trim=0 0 0 0,clip,width=\textwidth]{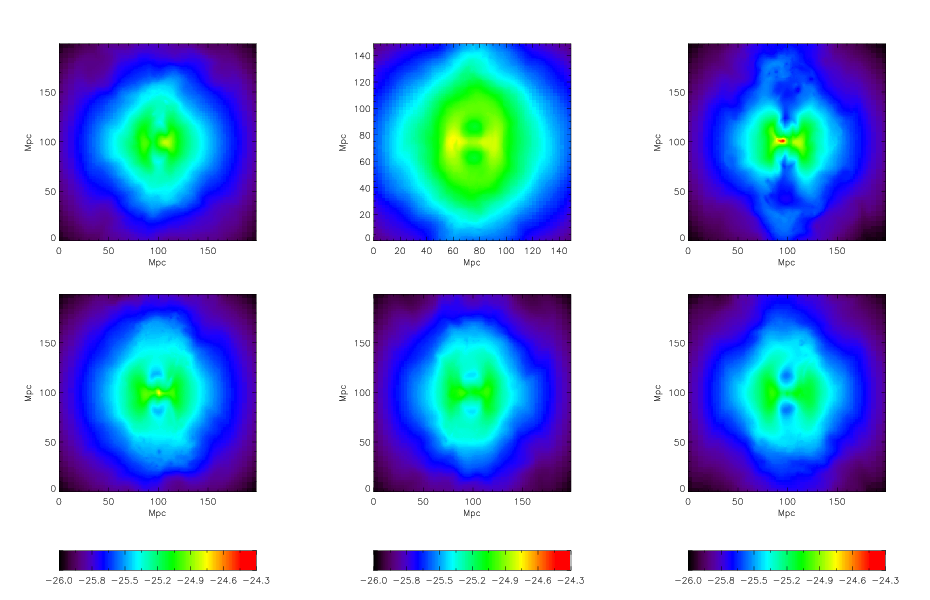}
\caption{Slices of the logarithm of the density at 1.5 Gyr. Top from left to right: D5-10, D5-8, D5-12. Bottom from left to right: D2-10, D10-10, D20-10.}
\label{fig:slices75}
\end{figure*} 

\begin{figure*}
\includegraphics[trim=0 0 0 0,clip,width=\textwidth]{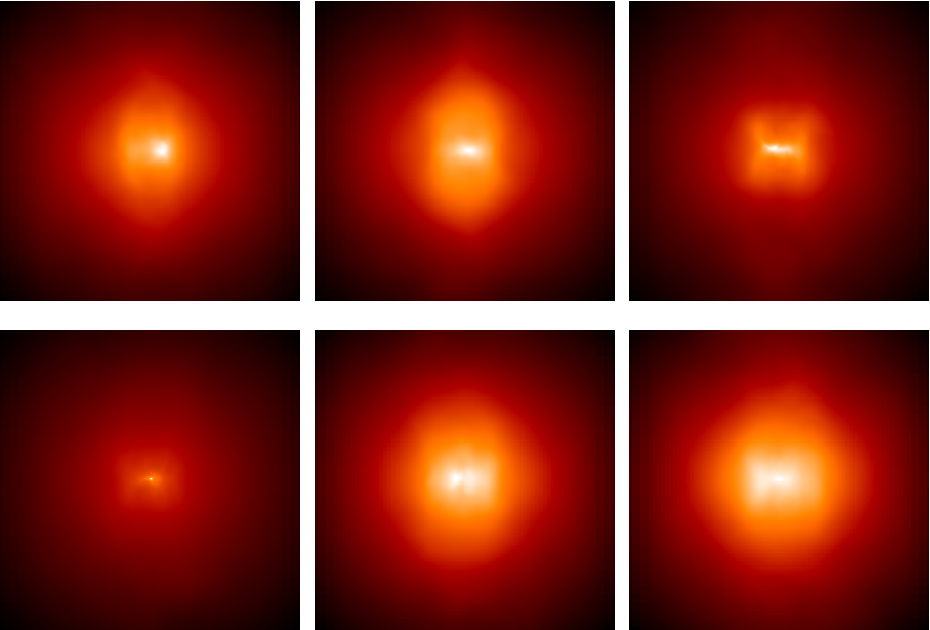}
\caption{X-ray maps (projections of $\Lambda(T) n_H n_e$) at 1 Gyr,  projected perpendicular to the bubble axis. Top from left to right: D5-10, D5-8, D5-12. Bottom from left to right: D2-10, D10-10, D20-10.}
\label{fig:xrayproj_lgscale_z_50}
\end{figure*} 

\begin{figure*}
\includegraphics[trim=0 0 0 0,clip,width=\textwidth]{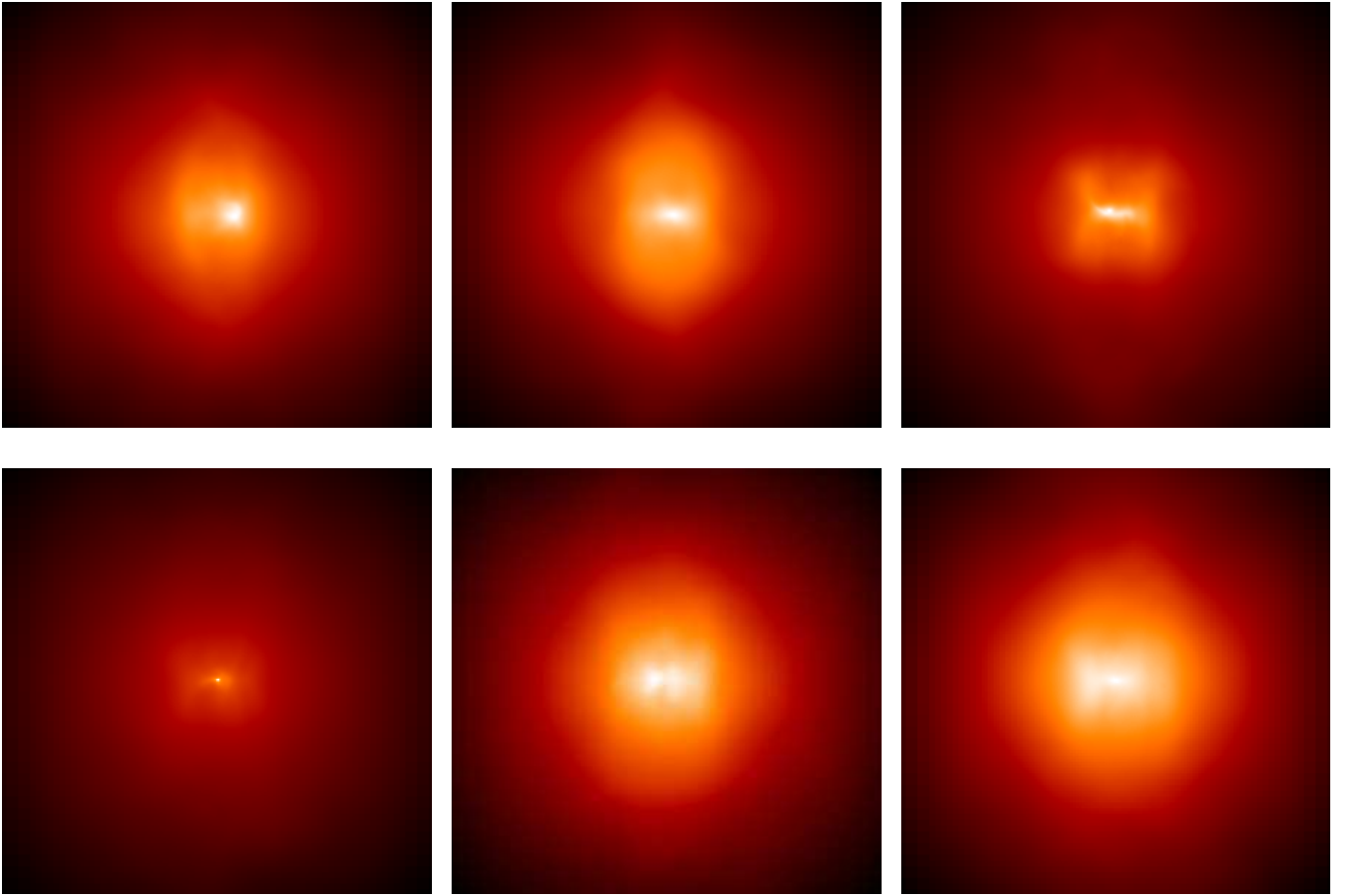}
\caption{X-ray maps (projections of $\Lambda(T) n_H n_e$) at 1.5 Gyr, projected perpendicular to the bubble axis. Top from left to right: D5-10, D5-8, D5-12. Bottom from left to right: D2-10, D10-10, D20-10.}
\label{fig:xrayproj_lgscale_z_75}
\end{figure*} 

In Fig.~\ref{fig:power2}, we vary the efficiency with which inflowing rest 
mass energy is converted into thermal energy deposited by the AGN. The
heating/cooling curves for runs D2-10, D5-10 and D10-10 look
remarkably similar. Even with different efficiencies, the AGN easily
regulates itself and the resulting duty cycles are largely unchanged.

In fact, the only run that shows a noticeably different evolution history is the high
energy D20-10 case.  Although this case shows a strong AGN event
after about 400 Mrs as in the other runs, unlike the other runs the
energy injection remains relatively weak at late times and varies irregularly.
So, provided the energy is injected sufficiently far from the region
that determines the fuelling of the AGN and provided the heating rate
is not so high as to destroy the cool core immediately, the duty cycle is
independent of the exact choice of parameters.

\subsection{Overall Cluster Properties}

Slices of the density through the cluster centre at $t=1$ Gyr and $t=1.5$ Gyr are shown in Fig. \ref{fig:slices50} - \ref{fig:slices75}, all plotted  on the same colour scale.  In these figures, the top rows show results from runs with varying bubble sizes at a fixed feedback efficiency,   and  the bottom rows show results from runs with varying efficiencies at a fixed bubble size. The $t=1$ Gyr slices shown in Fig.\ \ref{fig:slices50}  correspond to a time shortly after a major outburst has taken place in each of the runs.   One can see  large plumes streaming from the centre in all cases, although  these are somewhat weaker in the D5-8 run than in the other runs.   In general, the size of the energy injection region has a stronger impact on the appearance of the cluster than the choice of feedback efficiency, as one would expect for a self-regulating cycle, in which feedback increases only to the level necessary to balance cooling.
In fact, for efficiencies that vary by a factor of 10,  the density distributions look remarkably similar.  In the $t=1.5$ Gyr slices, shown in Fig.\ \ref{fig:slices75},   one only sees traces of the AGN activity in the form of bubbles of radius $\approx 10$ kpc on either side of the centre.  Again these bubbles are clearly seen in all runs, and the variance between runs is more strongly dependent on geometry than feedback efficiency.
 
X-ray maps corresponding to the same times as in the density slices are shown in Figs.~\ref{fig:xrayproj_lgscale_z_50} and \ref{fig:xrayproj_lgscale_z_75}.  For simplicity we construct these by projecting the full cooling emissivity as computed from $\epsilon =  n_H n_e \Lambda(T)$ as described in \S 2.2.   Because the emissivity is proportional to the square of the density, the X-ray  images are dominated by the emission from the cluster centre, and demonstrate the same overall peaked distribution observed in nearby cool-core clusters \citep[\eg][]{sarazin:86}. On closer inspection, one can also see substructure and signatures of the AGN-blown cavities. In the 1 Gyr profiles, for example, many runs show clear evidence of circular cavities that are surrounded by brighter emission, especially on the sides. This is particularly clear in run D5-12, which at 1 Gyr contains a large X-shaped structure that resembles the X-ray image of the Virgo elliptical M84 \citep{finoguenov:08}.  In all the runs, and  unlike in all previous attempts to simulate self-regulating AGN, the appearance of our self-regulated AGN in X-rays resembles observations of nearby clusters \citep[\eg][]{fabian:03, mcnamara:05, nulsen:05, forman:07, kraft:07}.

\begin{figure*}
\includegraphics[trim=0 0 0 0,clip,width=\textwidth]{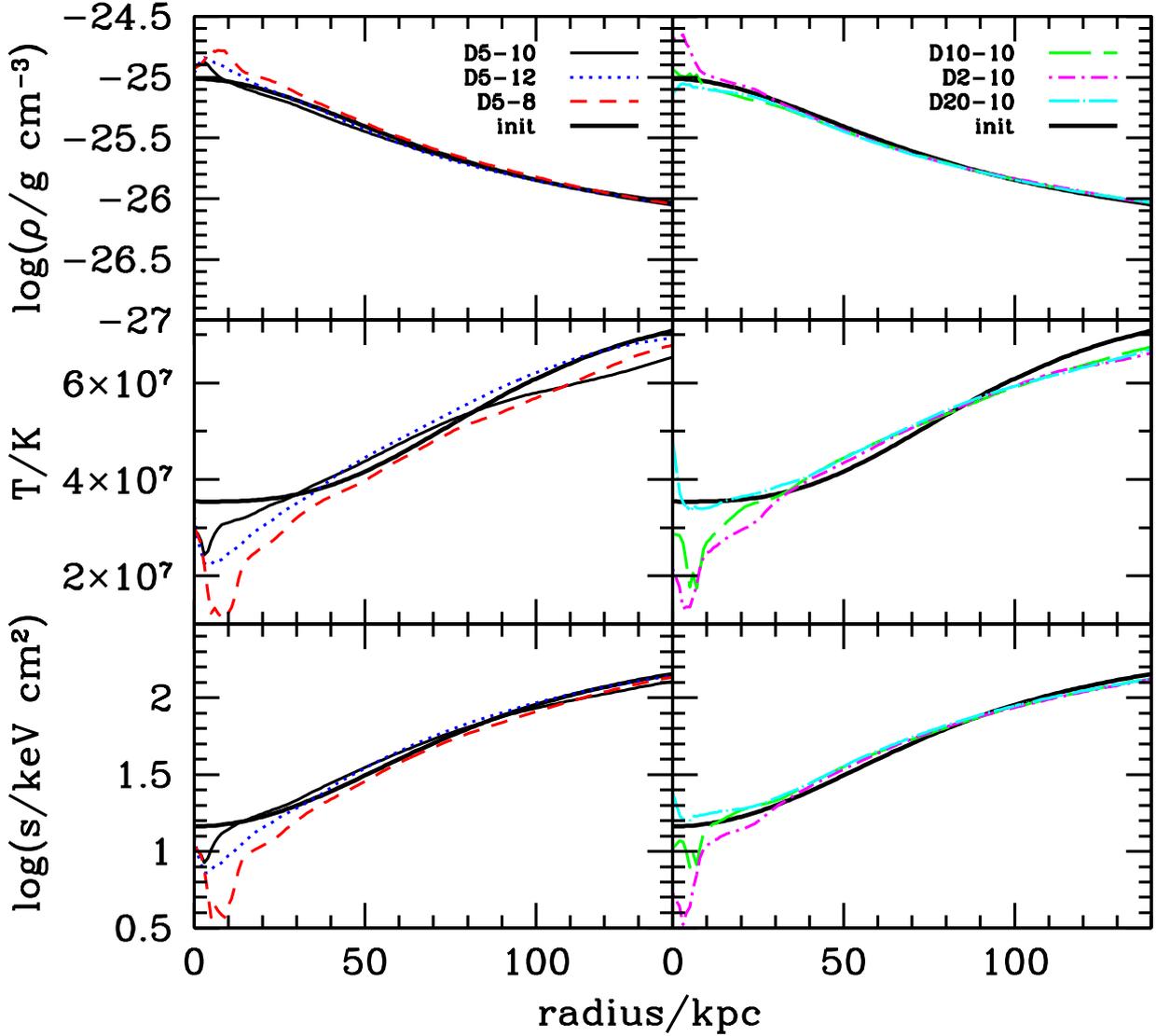}
\caption{Profiles of density (top), entropy (middle) and emission-weighted temperature (bottom) at 1.0 Gyr. In all panels the fat solid line shows the initial profile. Left panels: solid: D5-10; dotted: D5-12 and short dashed: D5-8. Right panels: long dashed: D10-10; dot-short dashed: D2-10 and dot-long dashed: D20-10.}
\label{fig:profiles50}
\end{figure*}

\begin{figure*}
\includegraphics[trim=0 0 0 0,clip,width=\textwidth]{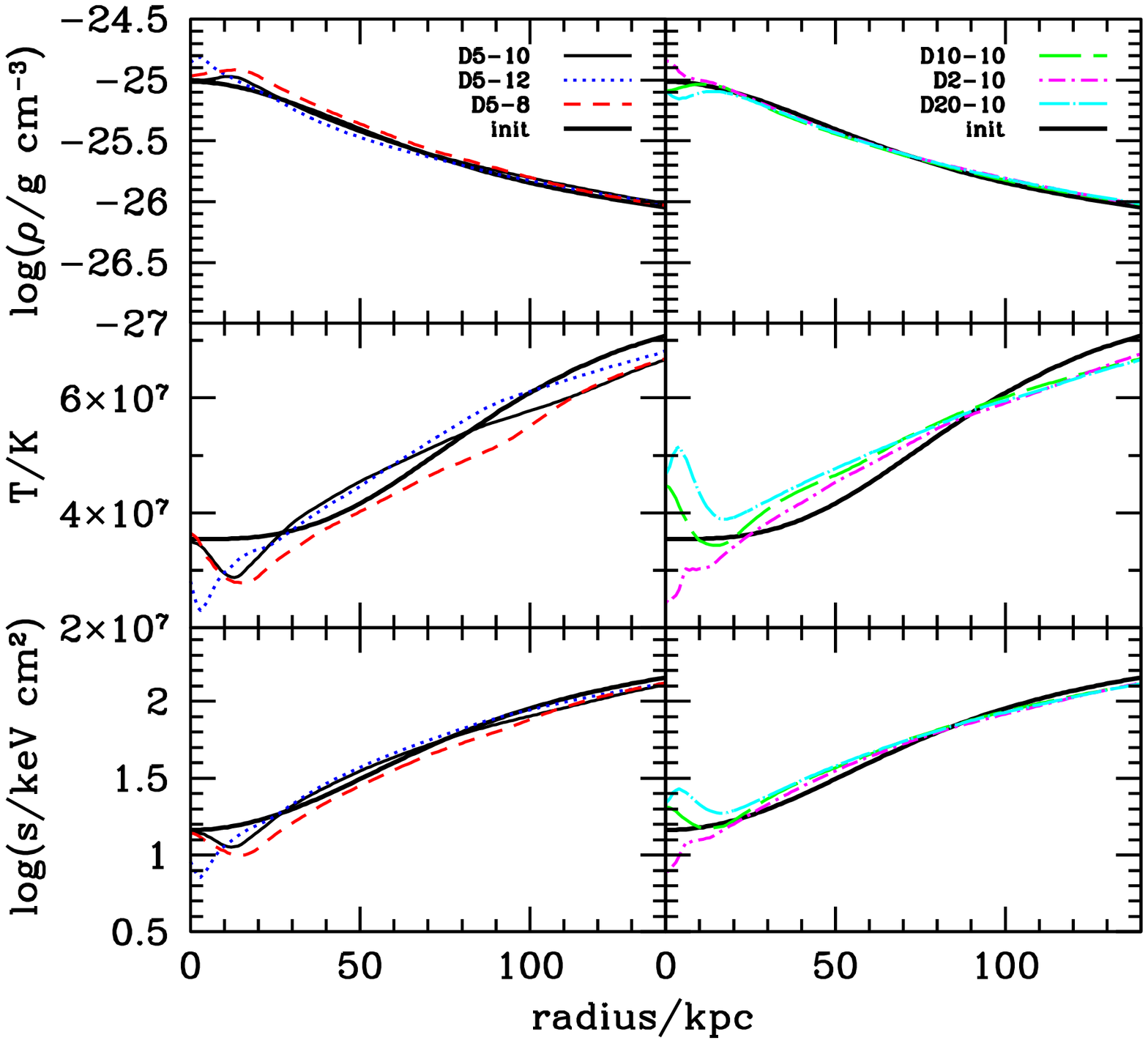}
\caption{Profiles of density (top), entropy (middle) and emission-weighted temperature (bottom) at 1.5 Gyr. In all panels the fat solid line shows the initial profile. Left panels: solid: D5-10; dotted: D5-12 and short dashed: D5-8. Right panels: long dashed: D10-10; dot-short dashed: D2-10 and dot-long dashed: D20-10.}
\label{fig:profiles75}
\end{figure*}

The radial profiles of average density, emission-weighted temperature, and entropy
are shown in Figs.~\ref{fig:profiles50} - \ref{fig:profiles75}.  
Here  the density is given by an average over the solid angle, $\Omega$, as
\begin{equation}
\bar \rho(r) = \frac{1}{4 \pi} \int d\Omega \, \bar \rho(r,\Omega),
\end{equation}
the emission-weighted temperature is given by
\begin{equation}
T_{\rm EW}(r)= \frac{\int d\Omega \, \epsilon(r,\Omega) \, T(r,\Omega)}{\int d\Omega \, \epsilon(r,\Omega)},
\end{equation}
and the radial entropy profiles is calculated from the emission-weighted temperature according to
\begin{equation}
S(r) = k_{\rm B}T_{\rm EW}(r) n_{\rm e}(r)^{-2/3} .
\end{equation}
In these figures, the right columns show the runs with
varying efficiencies, D2-10, D5-10 and D10-10, and the left columns
show the runs with different injection geometries, D5-8, D5-10 and
D5-12. 

Across the runs and at both time steps, the density contours deviate very little from the initial profile, as  shown by the thick solid lines.  On the other hand, there are noticeable changes in the temperature, which are also reflected in the entropy profiles.   These differences are largest within 50 kpc, and at these distances, the profiles at 1 Gyr show that 
significant cooling is going on at the cluster centre.   The clusters are caught in the midst of ongoing feedback
oscillations, and because these oscillations are not in phase with each other,  $T_{\rm EW}(R)$ and $S(R)$ at these small radii vary significantly across runs.  

Outside of 50 kpc, more subtle changes in temperature and entropy can be seen.  In the left panels, in which the
geometry of the bubbles is varied, the D5-8 run leads to slightly lower temperature and entropy profiles, the fiducial D5-10
stays close to the original profile, and the D5-12 run leads to a slight increase in temperature and entropy.   The cluster as a whole
is finding a slightly different equilibrium, which is weakly dependent on the geometry of energy input.  On the other hand,
in the right panels, in which the feedback efficiency is varied, all runs maintain $T_{\rm EW}(R)$ and $S(R)$ profiles
which are indistinguishable from each other.  Again this similarity is indicative of self-regulation and  occurs in spite of an order of magnitude variation in
feedback efficiency.

By $t$ = 1.5 Gyr, the central oscillations have died down and the central entropies have been raised to values
$\geq 10$ keV cm$^2$.  Outside of this core region, the profiles have evolved slightly in the D5-8 and D5-12 
cases, and are very similar to the initial state for all runs with $r_{\rm bubble} = 10$ kpc.  As is observed in cool-core clusters, the AGN has managed to maintain the original entropy propfile in the face of widespread cooling.  As we start already with a cluster profile that has non-zero core entropy, we cannot say much about the origin of the entropy floor in non-cool-core clusters \citep{cavagnolo:09}. However, these simulations, that cover a long physical time and many AGN outbursts, suggest that it is difficult to boost the core entropy to values  $\approx 100$ keV cm$^2$ with a string of relatively weak outbursts. 

\begin{figure*}
\includegraphics[trim=0 0 0 0,clip,width=\textwidth]{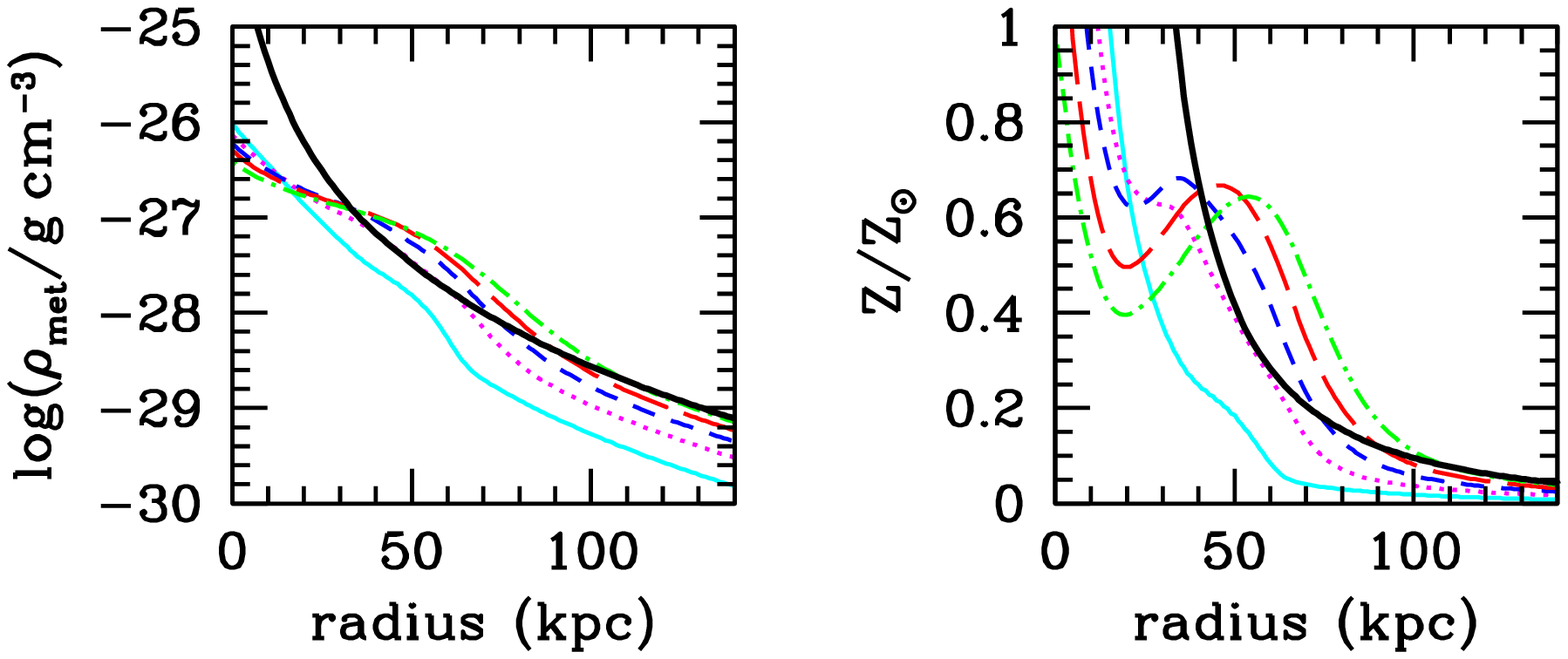}
\caption{Left: metal density of our fiducial run D5-10 at times t=400, 800, 1200, 1600 and 2000 Myr. The sold black line shows the expected metal density after 2 Gyr in the absence of any AGN activity. Right: The corresponding metallicity in units of solar metallicity (50 times metal mass fraction). Note that times 1200, 1600 and 2000 Myr show metallicity dips near the centre.}
\label{fig:metals}
\end{figure*}

Finally, in Fig.~\ref{fig:metals} we plot the metallicity in our simulated clusters. While previous simulations have modeled the effect of AGN activity on metals profiles \citep[\eg][]{tornatore:04, bruggen:02b}, none of these have been  self-regulating.  As a result of the intermittent AGN activity, the metals are distributed over a large volume.  Both metal density and metallicity are plotted in Fig.~\ref{fig:metals}, which shows that the AGN is effective at transporting metals.

A secondary maximum develops in the metallicity, which moves outward at around 20 km/s. As a result, the profiles of the metallicity at times 1200, 1600 and 2000 Myr show dips near the centre. This might explain the abundance dips in the iron distribution that have been observed in the Perseus and Centaurus clusters \citep[\eg][ and for a discussion on metal abundance patterns in clusters also see Simionescu et al. 2008]{matsushita:07, sanders:04}. In Perseus, the iron abundance dips from above 0.6 solar at around 40 kpc to below 0.5 solar near 15 kpc from the centre, and these numbers are similar to what we find here. However, unlike Perseus, the metal abundance in our simulations increases again in the very centre of the cluster, although this would not occur if the metal production had slowed down over time \citep{renzini:93}. Future simulations with a time-dependent metal production rate might shed further light on the origin of metallicity dips in some galaxy clusters.

The profiles of the turbulent parameters, $L$ and $K$, which we have not plotted here,  look very similar across runs. Over most of the cluster, the turbulent velocities are $\leq 100$ km s$^{-1},$ and the turbulent kinetic energy peaks at a radius $\approx 20$ kpc, at the outer edge of the bubbles where the turbulence is driven. The magnitude of turbulent velocities is in line with studies of resonant scattering in the ICM that find that the isotropic turbulent velocities on spatial 
scales smaller than 1 kpc are less than 100 km/s (Werner et al. 2009).

The effective turbulent diffusivities produced by the intermittent AGN are of the order of 500 kpc km/s \citep{rebusco:05}. \cite{david:08} have probed the Fe distribution in eight cool-core clusters and find that the Fe is significantly more extended than the stellar mass in the central galaxy. They find that diffusion coefficients of 300-3000 kpc km/s are required to produce the observed Fe profiles. Based on these inferred diffusion coefficients, they conclude that heating by turbulent diffusion of entropy and turbulent dissipation can balance the radiative losses in their sample. Our simulations confirm this. In real clusters, other factors  such as mergers may contribute to the broadening of the metallicity distribution. However, the magnitude of metal transport in galaxy clusters can be explained by the action of AGN alone.

\section{Discussion}

In this study, we have developed a complete and self-consistent three-dimensional model of AGN self-regulation in cool
core clusters.  In this picture cooling is balanced by periodic energy input from buoyant radio bubbles,
whose turbulent evolution couples them to the inflowing cool gas,  through mixing on scales that presently can
only be captured with subgrid modeling.   When turbulence is property accounted for, our model manages to regulate itself for a range of geometries and feedback efficiencies. A phase of episodic heating is followed by a phase in which the AGN is quiet as  the cool core reforms, which is followed in turn by another phase of episodic heating.  
Turbulence also helps to isotropise the injected energy over a turbulent timescale and remedies the problem of channel formation reported by \cite{vernaleo:06}.  Our results reproduce a number of key features seen in X-ray observations including:
\begin{itemize}
\item temperature, density and entropy profiles that stay fairly monotonous over time, despite rapid cooling near the cluster centre;
\item X-rays profiles that are strongly peaked towards the core, and show clear evidence of large bubble cavities rising towards the cluster outskirts; and
\item metals that are well mixed into the outer regions of the cluster and  exhibiting metallicity ``dips" near the cluster centre.
\end{itemize}

Despite these successes, however, there are a number of areas in which our model remains incomplete.
In our approach, feedback is controlled by the total mass inflow through a sphere of $r=3$ kpc, ignoring the details of the inner accretion flow, which will go through various phases as revealed by optical observations of  filaments \citep[\eg][]{fabian:08}.  Furthermore, our simulations do not include radiative heating, star formation, thermal conduction, cosmic rays,  or magnetic fields.   However, it is unclear how important these processes are for AGN self-regulation.
While thermal conduction may be important in establishing the self-similar entropy profiles found outside cluster cores \citep{cavagnolo:09}, its role in the cool cores of clusters is uncertain. Similarly, while cosmic rays are unlikely to supply much pressure support in galaxy clusters,  cosmic-ray pressure gradients lead to convection, which in turn heats the ICM by advecting internal energy inward and allowing the cosmic rays to do $p\,dV$ work on the thermal plasma \citep{chandran:07}.
Recently, there has also been renewed interest in the effect of magnetic fields on the dynamics of the ICM, which is subject to a number of instabilities including the magnetothermal instability \citep{balbus:00, parrish:08b} and the heat-flux driven buoyancy instability \citep{parrish:08a}.  These instabilities can lead to large-scale convective motions under a wide range of conditions. However, as demonstrated by numerical simulations \citep{parrish:08a}, these motions tend to shut themselves off and thus do not  play a major role in cool-core heating. Nonetheless, in future simulations, each of these effects should be explored.

The evolution of AGN in clusters involves an enormous range of time-scales, from the few days characteristic of the dynamical time at the event horizon to the few gigayears required for the development of a cooling catastrophe in the cluster core.  Here we advance a new explanation for the physics that determines the frequency of the intermittent bubbles that regulate cool-core clusters. Turbulence is crucial for mediating the heat input by AGN to the region that fuels the supermassive black hole, and thus the time scale for the process is determined by the turbulent turn-over time. This time scale, which is around a few $10^7$ years in our simulations, is given by  the cluster scale radius divided by the central sound speed, and thus depends on overall cluster properties rather than the details of energy injection.

A number of observational methods have been employed to quantify how long an average radio source spends in an active state and the time between active phases. One method invokes spectral ageing and lobe expansion speed arguments \citep{alexander:87}, yielding times of a few $10^7$ years.  Energy injection rates required to quench cooling flows point to time-scales of the order of few $10^7$ to $10^8$ years \citep[\eg][]{owen:98, mcnamara:05, nulsen:05}. Observations of multiple cavities in the Perseus and Virgo clusters also suggest times $\approx 10^8$ years between subsequent outbursts of the radio-loud AGN. Low-frequency radio surveys in clusters might also reveal radio ghosts whose synchrotron ages will provide important clues about the histories of  self-regulating radio-loud AGN.


\section*{Acknowledgements}

We thank the referee, Eugene Churazov, for helpful comments. 
MB acknowledges the support by the DFG grant BR 2026/3 within the Priority
Programme ``Witnesses of Cosmic History'' and the supercomputing grants NIC
2195 and 2256 at the John-Neumann Institut at the Forschungszentrum J\"ulich.
All simulations were conducted on the ÒSaguaroÓ cluster operated by the 
Fulton School of Engineering at Arizona State University.
The results presented were produced using the FLASH code, a product of the DOE
ASC/Alliances-funded Center for Astrophysical Thermonuclear Flashes at the
University of Chicago.


%
\bibliographystyle{mn2e}
\bibliography{%
feedback,%
bubble_evol%
}

\bsp

\label{lastpage}

\end{document}